\documentclass[altaffilletter,superscriptaddress,amsmath,amssymb,twocolumn]{revtex4-1}
\usepackage{xcolor}
\usepackage{physics}
\usepackage{graphicx}
\usepackage{dcolumn}
\usepackage{bm}

\begin{document}

\title[]{Coupled-waveguides for dispersion compensation in semiconductor lasers}
\affiliation{Institute for Quantum Electronics, ETH-Zurich, CH-8093 Zurich, Switzerland}
\author{Yves Bidaux}
\email{bidauxy@phys.ethz.ch}
\author{Filippos Kapsalidis}
\author{Pierre Jouy}
\author{Mattias Beck}
\author{J\'er\^ome Faist}
\email{jfaist@ethz.ch}
\date{\today}

\maketitle

\textbf{The generation of optical frequency combs via direct electrical pumping of semiconductor lasers~\cite{derickson_short_1992,Avrutin_monolithic_2000,keller_passively_2006,rafailov_mode-locked_2007,hugi_mid-infrared_2012} is an attractive alternative to the well-established mode-locked laser sources in terms of compactness, robustness and integrability. However, the high chromatic dispersion of bulk semiconductor materials~\cite{palik_handbook_1998} can prevent the generation of frequency combs~\cite{villares_dispersion_2016} or lead to undesired pulse lengthening~\cite{kim_pulse_2006}. In this letter, we present a novel dual waveguide for intracavity dispersion compensation in semiconductor lasers. We apply the concept to a short mid-infrared wavelength quantum cascade laser operating in the first atmospheric window ($\lambda \approx$ 4.6 $\mu$m). As a result, stable comb operation on the full dynamical range is achieved in this device. Unlike previously proposed schemes, the dual waveguide approach can be applied to various types of semiconductor lasers and material systems. In particular, it could enable efficient group velocity dispersion compensation at near-infrared wavelengths where semiconductor materials exhibit a large value of that parameter.}

Recently, novel sensing methods based on optical frequency combs~\cite{diddams_evolving_2010} have attracted much attention. In particular, the dual comb configuration offers, for spectroscopy, the key advantage of combining high resolution with large spectral coverage while not requiring any moving part~\cite{keilmann_time-domain_2004}. Quantum cascade lasers (QCLs) can operate as optical frequency combs~\cite{hugi_mid-infrared_2012,rosch_octave-spanning_2014,burghoff_terahertz_2014} enabling the realization of compact mid-infrared dual-comb spectrometers~\cite{villares_dual-comb_2014,yang_terahertz_2016,westberg_mid-infrared_2017}. The signal over noise characteristics of dual comb spectrometers heavily depend on the linewidth and stability of the comb devices. Recent experiments where Gires-Tournois dielectric coatings were deposited on the back facet of QCL devices showed the essential role of dispersion compensation for the improvement of the combs bandwidth and operation stability~\cite{villares_dispersion_2016,lu_dispersion_2017}. However, dispersion compensation by waveguide engineering~\cite{riemensberger_dispersion_2012,okawachi_bandwidth_2014} is preferable in terms of its capability to correct for large amounts of dispersion, as well as its reliability and cost.

In a coupled-waveguide system, the wavelength dependence of the inter-waveguide coupling through the evanescent tails of the individual modes may lead to a significant change of its propagation vector. In turn, the modes of coupled-waveguides can exhibit an additional dispersion that exceeds the one of bulk materials~\cite{peschel_compact_1995}. In a previous work, we found that the coupling of the laser fundamental mode to a plasmonic resonance engineered in the lasers top cladding can lead to efficient dispersion compensation in long ($\lambda$ $\approx$ 7 $\mu$m) mid-infrared wavelength QCLs~\cite{bidaux_plasmon-enhanced_2017,jouy_dual_2017}. This approach cannot be applied to shorter wavelengths where dispersion is particularly high due to the proximity of the bandgap~\cite{bidaux_investigation_2017}. This is due to the difficulty to scale the plasma frequency accordingly (see Supplementary Fig.~1) and to the introduction of unavoidable free carrier losses.

\begin{figure*}[]
\centering
\includegraphics[width=0.95\textwidth]{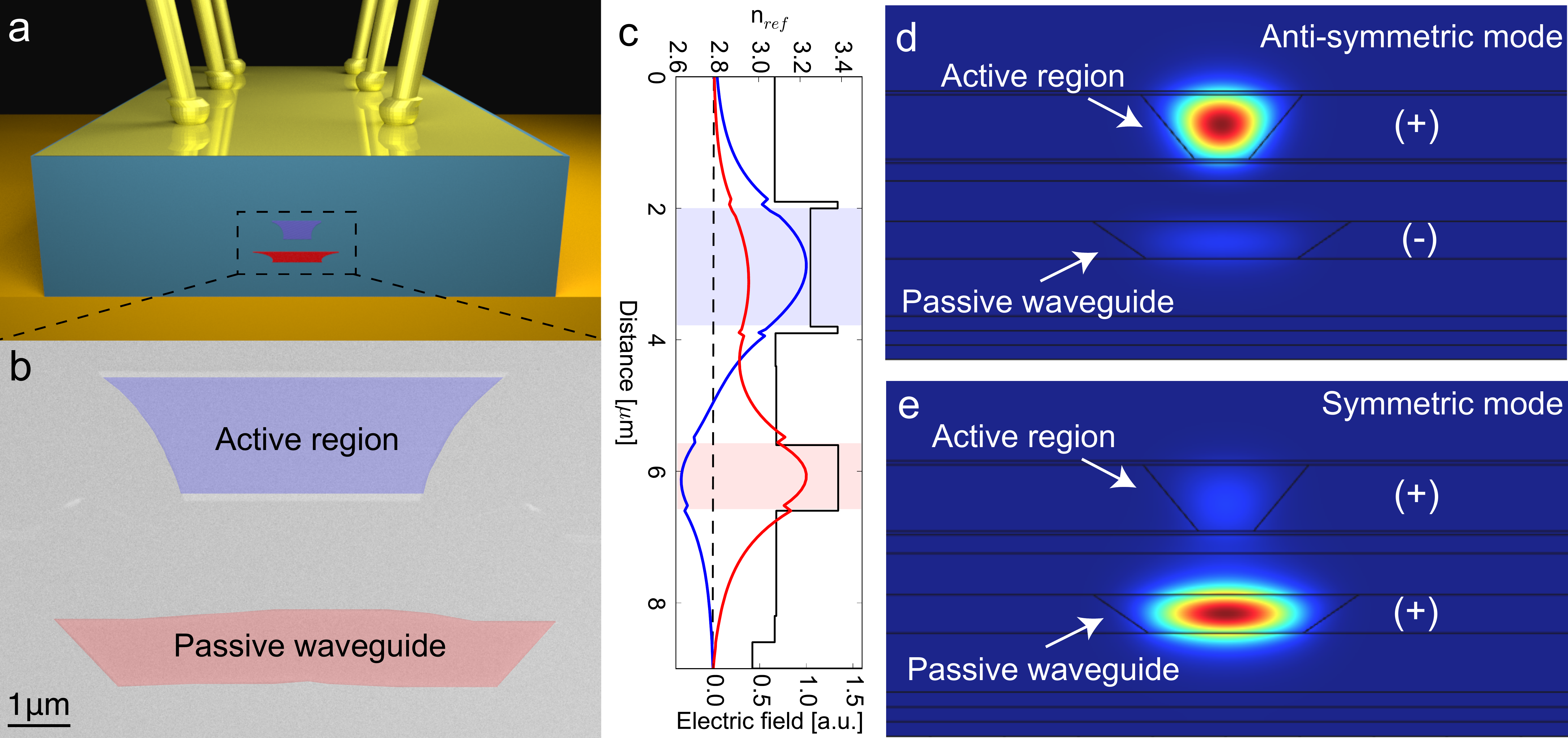}
\caption{\textbf{Device cross-section.} \textbf{a}, Schematic view of a mounted device. \textbf{b}, SEM picture of the fabricated device. The passive waveguide is coloured with red while the active region is coloured in blue. \textbf{c}, Cut in the vertical plane of the electric field profile of the anti-symmetric (blue line) and symmetric (red line) modes and refractive index profiles. \textbf{d}, Intensity profile of the anti-symmetric mode at 2220 cm$^{-1}$. \textbf{e}, Intensity profile of the symmetric mode at 2220 cm$^{-1}$.}
\label{modes}
\end{figure*}

\begin{figure}[]
\centering
\includegraphics[width=0.45\textwidth]{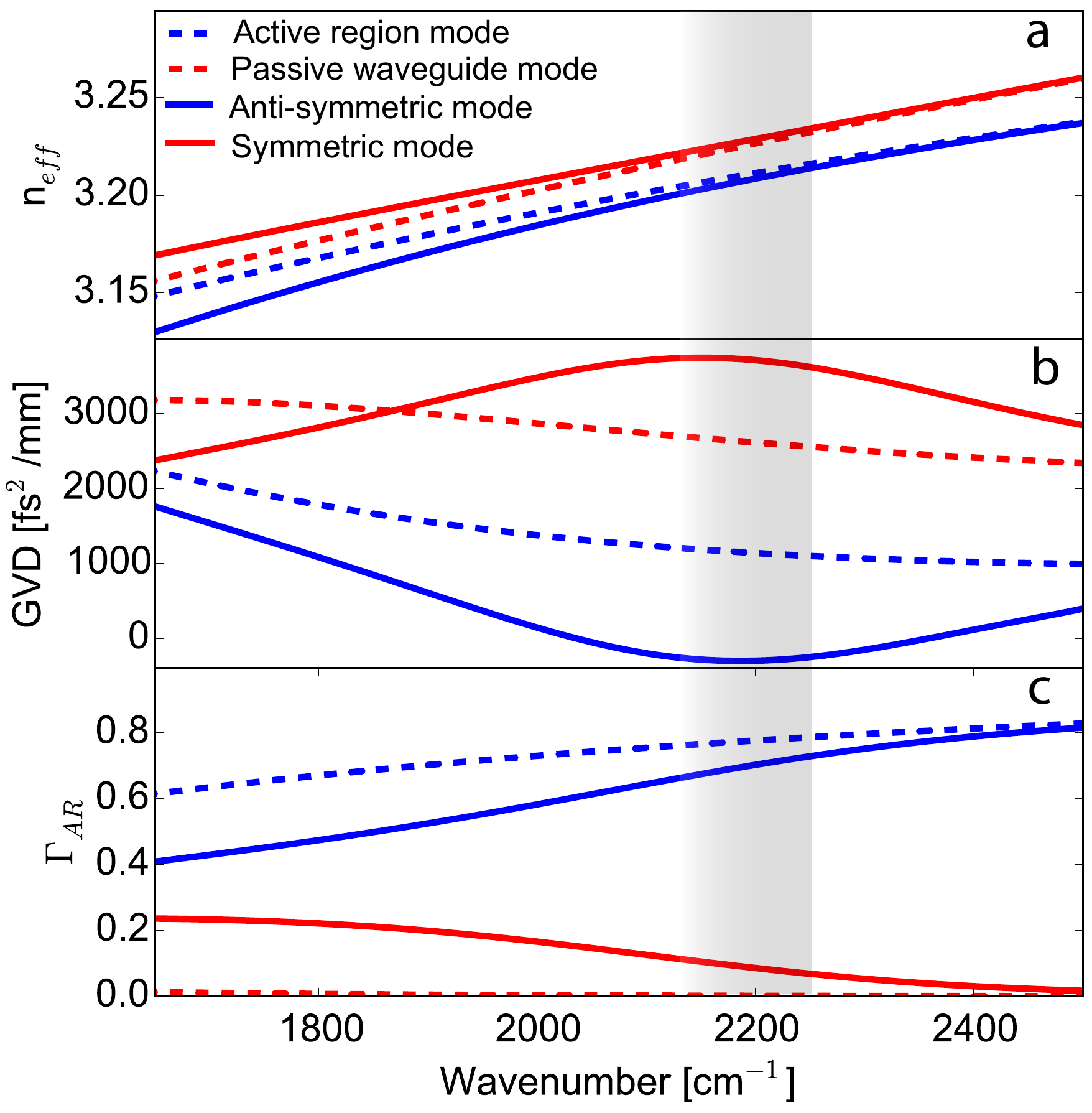}
\caption{\textbf{Simulation of the dispersion.} \textbf{a}, Effective refractive index, \textbf{b}, GVD as a function of frequency and \textbf{c}, overlap factor for the active region (dashed blue line), passive waveguide (dashed red line), anti-symmetric (blue line) and symmetric (red line) modes.}
\label{dispersion}
\end{figure}
In the present work, we demonstrate dispersion compensation by coupling the optical mode to a passive dielectric waveguide engineered in the device top cladding. We fabricated a QCL based on a diagonal three quantum well design for emission at 4.6 $\mu$m~\cite{yang_onhigh-peak-power_2008,bismuto_influence_2011}. As seen in Fig.~\ref{modes}a, the device integrates a passive InGaAs waveguide close to the active region (see Methods for fabrication details). A Scanning Electron Micrograph (SEM) image of the device front facet is displayed in Fig.~\ref{modes}b. Naturally, each waveguide has a frequency-dependent propagation vector. The waveguides are designed such that these two propagation vectors are equal for a resonant frequency $\omega_0$. At this frequency, if the two optical modes have equal group velocity dispersion (GVD), these couple to form a symmetric and an anti-symmetric supermode with resonantly depressed (respectively enhanced) GVD that can then be used for dispersion compensation~\cite{peschel_compact_1995}. Applying directly such approach to a laser system leads to a fundamental difficulty: Since at resonance the supermodes are equally distributed in both waveguides, lasing can occur on the high and low GVD supermodes simultaneously.

We therefore considered a system where the two modes of the individual waveguides have a different GVD. In that case, using coupled mode theory, the induced GVD ($\Delta$GVD) writes as (see Supplementary materials for the derivation):
\begin{equation}
\Delta \text{GVD}_{+/-} \simeq \pm\frac{\kappa}{\delta \omega^2} \left(\tilde{\omega}^2 +1 \right)^{-3/2} \left( 1 + \alpha \tilde{\omega}\right)
\end{equation}
for the symmetric (+) and anti-symmetric (-) modes, where $\kappa$  is the coupling strength, $\tilde{\omega} =\left( \omega -\omega_0\right) /\delta\omega$ is the normalised frequency, $\delta\omega = 2 \kappa  \abs{ \frac{1}{\nu_1} - \frac{1}{\nu_2} }  ^{-1}$ the resonance bandwidth and $\alpha = \frac{6 \kappa}{\delta \omega^2} \left(D_1 -D_2 \right)$ the asymmetry factor. In the last expressions, $\nu_i$ are the group velocities and $D_i$ the GVDs of the decoupled active region (i=1) and passive waveguide (i=2) modes. For $\alpha$ different from zero, the GVD difference between the two waveguides $D_1 -D_2$ induces a shift of the GVD maxima. In turn, dispersion compensation can be achieved without requiring that the two optical modes are in resonance. The resulting asymmetry in the overlap factor of the two supermodes with the active region enables the laser to naturally select one of the two and remain monomode.

We simulated the optical modes for various geometries using a finite element solver (see Methods) and targeted zero GVD at the central laser wavelength. The refractive index profile along the growth axis of the fabricated device is displayed in Fig.~\ref{modes}c. The computed electric field profile of the anti-symmetric (blue line) and symmetric (red line) modes are also displayed, while the full two dimensional intensity profiles of these modes are reported in Fig.~\ref{modes}d and e. These two modes result from the coupling of the two distinct active region and passive waveguide optical modes with GVD $D_1$ = 1140 fs$^2$/mm and $D_2$ = 2600 fs$^2$/mm. The dispersion curves of the two uncoupled modes are reported in Fig.~\ref{dispersion}a with dashed lines while the dispersion of the coupled modes are reported with full lines. Our simulation predicts a minimum of the GVD close to zero at the laser operation wavelength for the anti-symmetric mode (see Fig.~\ref{dispersion}b, shaded area) and an overlap with the active region equal to 0.7 (see Fig.~\ref{dispersion}c). The latter has as a result a much larger overlap with the active region ($\Gamma_{AR}$) than the symmetric mode which is localized in the passive waveguide. This induces a strong mode selection mechanism which ensures that the device operates with high efficiency on the low GVD anti-symmetric mode.

\begin{figure}[]
\centering
\includegraphics[width=0.48\textwidth]{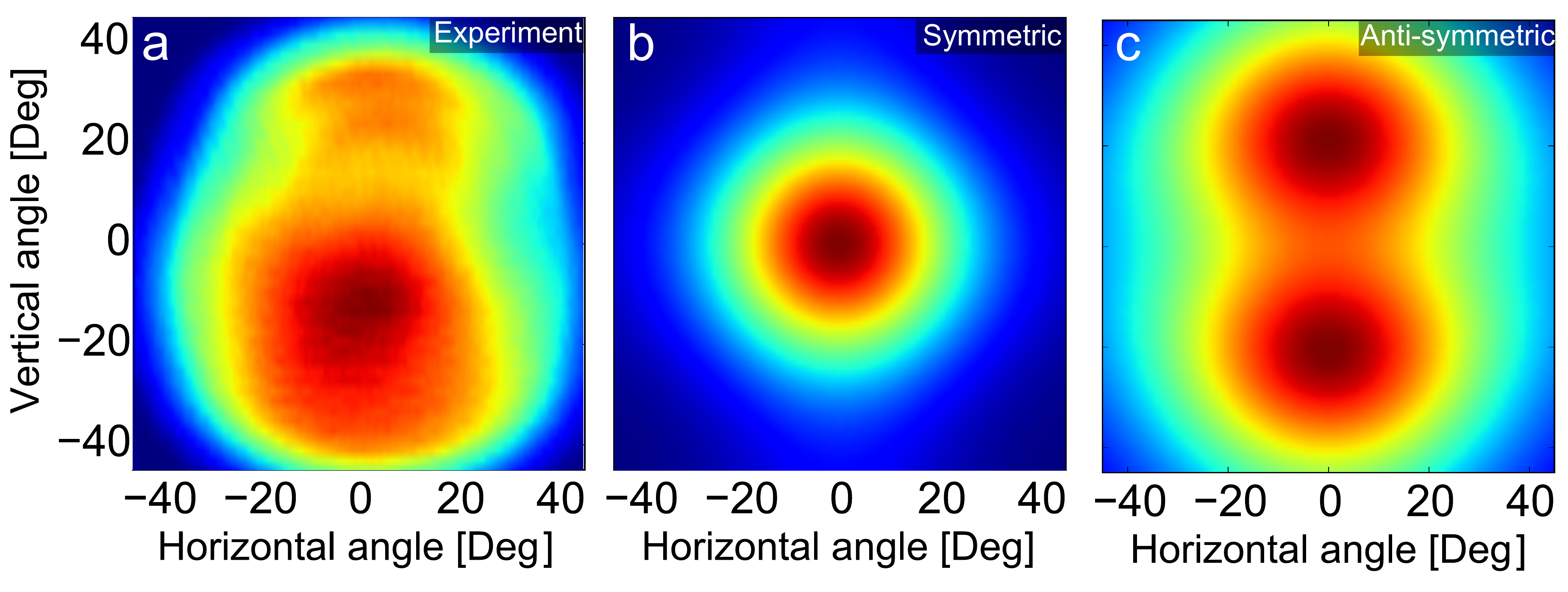}
\caption{\textbf{Farfield.} \textbf{a}, Farfield emission profile of the device measured in pulsed mode for a current of 830 mA. \textbf{b-c}, Simulated farfield emission profile for the symmetric and the anti-symmetric modes.}
\label{chara1}
\end{figure}

To confirm that the laser operates on the anti-symmetric mode, we measured its farfield emission profile in pulsed operation at rollover. These experimental data (see Fig.~\ref{chara1}a) are compared to the farfield computed from the supermodes nearfield obtained from our 2D mode simulation that are displayed in Fig.~\ref{chara1}b-c. The difference between the measured (Fig.~\ref{chara1}a) and computed (Fig.~\ref{chara1}c) farfield is attributed to the presence of loss and gain in the passive waveguide and active region, respectively, which is not taken into account by our model.

\begin{figure}[]
\centering
\includegraphics[width=0.48\textwidth]{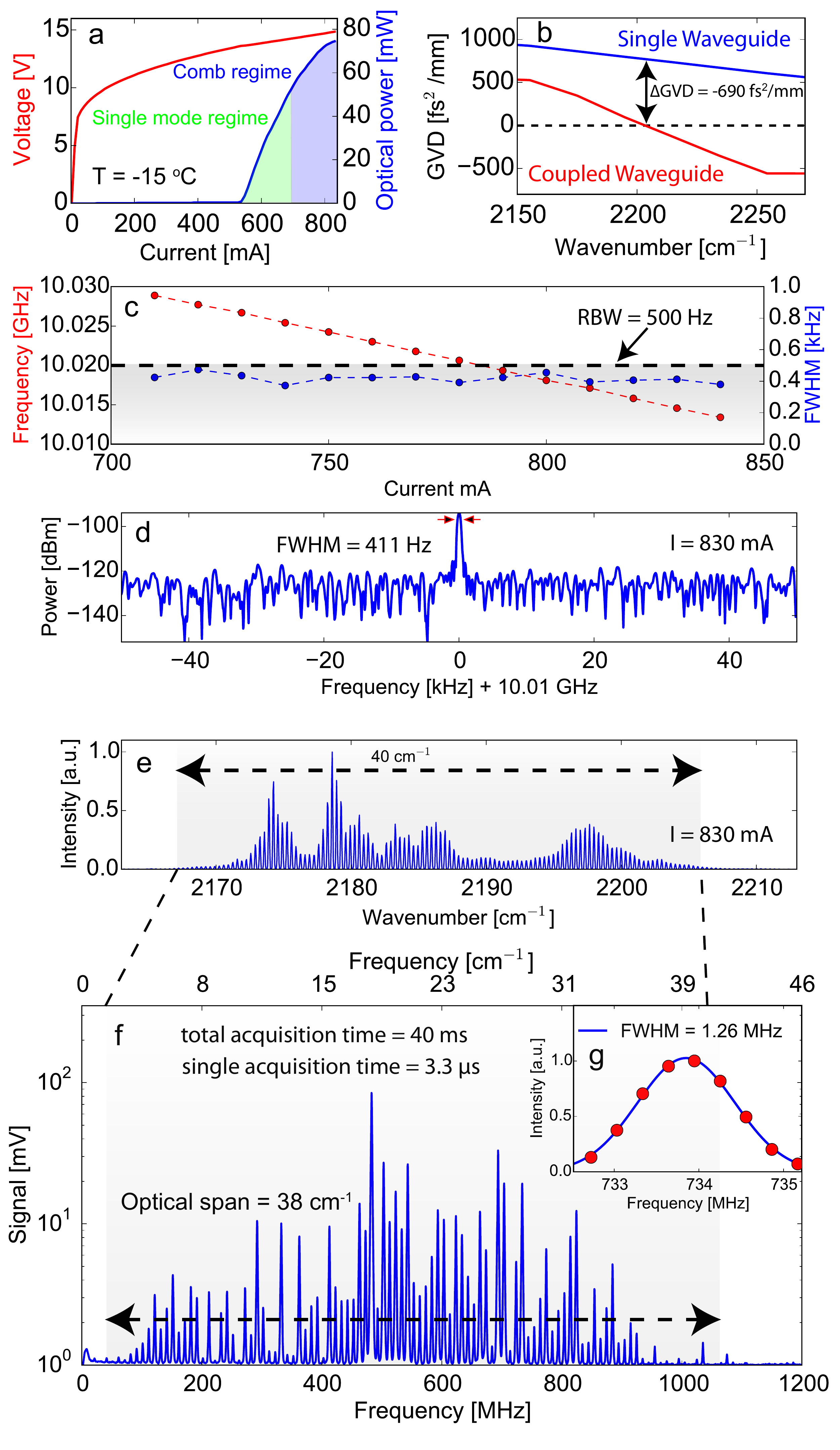}
\caption{\textbf{Frequency combs characterization.} \textbf{a}, Light and voltage versus current characteristics of the device measured at -15 $^{\circ}$C in continuous wave. \textbf{b}, GVD measured at -15 $^{\circ}$C for a current of 450 mA for the device (red line) and for a current of 265 mA for the reference device (blue line). \textbf{c}, Intermode beatnote frequency (blue) and FWHM (red) as a function of current measured at -15 $^{\circ}$C with a resolution bandwidth of 500 Hz. \textbf{d}, Intermode beatnote measured at -15 $^{\circ}$C for a current of 830 mA. \textbf{e}, Optical spectrum measured in continuous wave at -15 $^{\circ}$C for a current of 830 mA. \textbf{f}, Multiheterodyne beating spectrum of the two frequency combs. 93 distinct comb lines can be seen corresponding to an optical span of 38 cm$^{-1}$. \textbf{g}, The FWHM of one line is 1.26 MHz.}
\label{chara2}
\end{figure}

We measured the light and voltage versus current characteristics of the device while operated under continuous wave operation at -15 $^{\circ}$C. The output power reached 75 mW at rollover (see Fig.~\ref{chara2}a). We deduced the GVD from the subthreshold emission spectrum~\cite{hofstetter_measurement_1999} measured at a current of 450 mA at -15 $^{\circ}$C. The GVD is displayed in Fig.~\ref{chara2}b (red line). For comparison, a measurement performed on a 4.5 mm long reference device without the additional passive waveguide (see Methods for fabrication details and Supplementary Fig.~2 for full characterization) is also displayed (blue line). For the reference device, a GVD as large as 740 fs$^2$/mm is deduced at 2200 cm$^{-1}$, while for the device including the passive waveguide a GVD of only 50 fs$^2$/mm is deduced at 2200 cm$^{-1}$. This confirms that the coupling to the passive waveguide reduced the dispersion of the device ($\Delta$GVD = -690 fs$^2$/mm).

To characterize the coherence of the optical spectrum we measured the intermode beatnote of the device directly from the laser bias current using a RF spectrum analyzer~\cite{villares_dual-comb_2014} with a resolution bandwidth (RBW) of 500 Hz. Narrow beatnotes with full width at half maximum (FWHM) less than 500 Hz were measured on the whole dynamical range (see Fig.~\ref{chara2}c and Supplementary Fig.~3). The intermode beatnote frequency tuned smoothly with current at a rate of -0.13 GHz/A, similar to values reported in~\citet{bidaux_plasmon-enhanced_2017}. At a current of 830 mA the intermode beatnote has a FWHM $\simeq$ 411 Hz (Fig.~\ref{chara2}d) and the optical spectrum spans 40 cm$^{-1}$ (see Fig.~\ref{chara2}e). Such narrow and stable beatnotes have up to now always been observed in QCLs operating in comb regime.

Nevertheless, to demonstrate that the devices actually operate as frequency combs, we measured the multiheterodyne beating spectrum of two HR coated 4 mm devices at -20 $^{\circ}$C. The first device current was set to 600 mA while the second device current was set to 460 mA. The optical beams were combined and focused on a MCT detector and the signal was recorded using an oscilloscope. The total acquisition time was 40 ms and the signal was averaged with a single acquisition time of 3.3 $\mu$s using the method detailed in~\citet{villares_dual-comb_2014}. The resulting spectrum is displayed in Fig.~\ref{chara2}f. Ninety three distinct comb lines are observed. Since the free spectral range of the combs is approximately 0.39 cm$^{-1}$, this corresponds to an optical span of 38 cm$^{-1}$ and demonstrates that the devices operate as frequency combs on the whole spectral bandwidth. The FWHM of one line is 1.26 MHz (see Fig.~\ref{chara2}g), this directly shows the suitability of these devices for dual-comb spectrometers in the first atmospheric window.

In conclusion, we presented a dual coupled-waveguide design for dispersion compensation in semiconductor lasers and demonstrated its application to a quantum cascade laser. The fabricated devices operate as frequency combs on the whole spectral bandwidth and dynamical range. Unlike previously demonstrated integrated dispersion compensation methods, the dual waveguide scheme can be applied to various types of semiconductor lasers and material systems. Moreover, it allows to correct for GVD values up to a few thousand fs$^2$/mm. This suggests that this design strategy could enable the development of high performance semiconductor laser based optical frequency combs from the long mid-infrared to the near-infrared wavelengths ranges. Actually, an interband laser waveguide design that is computed to be dispersion compensated at 1.3 $\mu$m is proposed in the supplementary material.

\section*{Methods}

\subsection*{Devices fabrication}
The active region of the device including coupled-waveguides was grown by molecular beam epitaxy using the InGaAs/AlInAs material system on an InP:Si (2 x 10$^{17}$ cm$^{-3}$) substrate. The active layers were placed between two 100 nm thin layers of Si-doped InGaAs (6 x 10$^{16}$ cm$^{-3}$). The exact layers sequence of the QCL is detailed in~\citet{yang_onhigh-peak-power_2008}. The device was processed to a buried heterostructure QCL. A 4.4 $\mu$m wide ridge was etched by wet-etching technique and a lateral InP:Fe insulating layer was grown by metalorganic vapour phase epitaxy (MOVPE). The top cladding layers were then grown by MOVPE and include an InGaAs waveguiding layer. The top cladding layers sequence is 0.5 $\mu$m InP:Si 1 x 10$^{17}$ cm$^{-3}$, 1.2 $\mu$m InP:Si 1 x 10$^{16}$ cm$^{-3}$, 1 $\mu$m InGaAs:Si 1 x 10$^{16}$ cm$^{-3}$, 1.6 $\mu$m InP:Si 1 x 10$^{16}$ cm$^{-3}$, 0.4 $\mu$m InP:Si 2 x 10$^{17}$ cm$^{-3}$, 0.4 $\mu$m InP:Si 3 x 10$^{18}$ cm$^{-3}$. A second 6.7 $\mu$m wide ridge was etched in the cladding by wet-etching and a lateral InP:Fe insulating layer was grown by MOVPE. For the reference device the top cladding layers sequence is: 0.42 $\mu$m InP:Si 1 x 10$^{17}$ cm$^{-3}$, 2.5 $\mu$m InP:Si 1 x 10$^{16}$ cm$^{-3}$, 0.85 $\mu$m InP:Si 1 x 10$^{18}$ cm$^{-3}$. Devices were mounted episide down on AlN submounts.

\subsection*{Optical mode simulation}
We performed two-dimensional finite-element simulations of the waveguide structure using the commercially available software Comsol Multiphysics. The intrinsic refractive indices were obtained from Pikhtin and Yas'kov formula using the parameters given in~\citet{palik_handbook_1998}. The doped refractive indices are deduced using Drude model. We assume a collision time of 100 fs and effective masses of 0.073 m$_e$ for InP and 0.043 m$_e$ for InGaAs.

\section*{Acknowledgements}

The authors would like to thank Dmitry Kazakov and Matthew Singleton for careful reading of the paper and fruitful discussions. This work was supported by the Swiss National Science Fundation (SNF) and by DARPA under the SCOUT program.

\section*{Authors contributions}
Y.B. developed the simulations, designed the waveguide, performed experiments, analysed data and wrote the manuscript. F.K. fabricated the coupled-waveguide quantum cascade lasers, performed experiments and analysed data. P.J. fabricated the reference quantum cascade lasers, performed experiments and analysed data. M.B. grew the quantum cascade laser material. J.F. proposed the experiment, analysed data and supervised the work.


\begin{thebibliography}{25}%
\makeatletter
\providecommand \@ifxundefined [1]{%
 \@ifx{#1\undefined}
}%
\providecommand \@ifnum [1]{%
 \ifnum #1\expandafter \@firstoftwo
 \else \expandafter \@secondoftwo
 \fi
}%
\providecommand \@ifx [1]{%
 \ifx #1\expandafter \@firstoftwo
 \else \expandafter \@secondoftwo
 \fi
}%
\providecommand \natexlab [1]{#1}%
\providecommand \enquote  [1]{``#1''}%
\providecommand \bibnamefont  [1]{#1}%
\providecommand \bibfnamefont [1]{#1}%
\providecommand \citenamefont [1]{#1}%
\providecommand \href@noop [0]{\@secondoftwo}%
\providecommand \href [0]{\begingroup \@sanitize@url \@href}%
\providecommand \@href[1]{\@@startlink{#1}\@@href}%
\providecommand \@@href[1]{\endgroup#1\@@endlink}%
\providecommand \@sanitize@url [0]{\catcode `\\12\catcode `\$12\catcode
  `\&12\catcode `\#12\catcode `\^12\catcode `\_12\catcode `\%12\relax}%
\providecommand \@@startlink[1]{}%
\providecommand \@@endlink[0]{}%
\providecommand \url  [0]{\begingroup\@sanitize@url \@url }%
\providecommand \@url [1]{\endgroup\@href {#1}{\urlprefix }}%
\providecommand \urlprefix  [0]{URL }%
\providecommand \Eprint [0]{\href }%
\providecommand \doibase [0]{http://dx.doi.org/}%
\providecommand \selectlanguage [0]{\@gobble}%
\providecommand \bibinfo  [0]{\@secondoftwo}%
\providecommand \bibfield  [0]{\@secondoftwo}%
\providecommand \translation [1]{[#1]}%
\providecommand \BibitemOpen [0]{}%
\providecommand \bibitemStop [0]{}%
\providecommand \bibitemNoStop [0]{.\EOS\space}%
\providecommand \EOS [0]{\spacefactor3000\relax}%
\providecommand \BibitemShut  [1]{\csname bibitem#1\endcsname}%
\let\auto@bib@innerbib\@empty
%</preamble>
\bibitem [{\citenamefont {Derickson}\ \emph {et~al.}(1992)\citenamefont
  {Derickson}, \citenamefont {Helkey}, \citenamefont {Mar}, \citenamefont
  {Karin}, \citenamefont {Wasserbauer},\ and\ \citenamefont
  {Bowers}}]{derickson_short_1992}%
  \BibitemOpen
  \bibfield  {author} {\bibinfo {author} {\bibfnamefont {D.~J.}\ \bibnamefont
  {Derickson}}, \bibinfo {author} {\bibfnamefont {R.~J.}\ \bibnamefont
  {Helkey}}, \bibinfo {author} {\bibfnamefont {A.}~\bibnamefont {Mar}},
  \bibinfo {author} {\bibfnamefont {J.~R.}\ \bibnamefont {Karin}}, \bibinfo
  {author} {\bibfnamefont {J.~G.}\ \bibnamefont {Wasserbauer}}, \ and\ \bibinfo
  {author} {\bibfnamefont {J.~E.}\ \bibnamefont {Bowers}},\ }\bibfield  {title}
  {\enquote {\bibinfo {title} {Short pulse generation using multisegment
  mode-locked semiconductor lasers},}\ }\href@noop {} {\bibfield  {journal}
  {\bibinfo  {journal} {IEEE Journal of Quantum Electronics}\ }\textbf
  {\bibinfo {volume} {28}},\ \bibinfo {pages} {2186--2202} (\bibinfo {year}
  {1992})}\BibitemShut {NoStop}%
\bibitem [{\citenamefont {Avrutin}, \citenamefont {Marsh},\ and\ \citenamefont
  {Portnoi}(2000)}]{Avrutin_monolithic_2000}%
  \BibitemOpen
  \bibfield  {author} {\bibinfo {author} {\bibfnamefont {E.}~\bibnamefont
  {Avrutin}}, \bibinfo {author} {\bibfnamefont {J.}~\bibnamefont {Marsh}}, \
  and\ \bibinfo {author} {\bibfnamefont {E.}~\bibnamefont {Portnoi}},\
  }\bibfield  {title} {\enquote {\bibinfo {title} {Monolithic and
  multi-{GigaHertz} mode-locked semiconductor lasers: {Constructions},
  experiments, models and applications},}\ }\href {\doibase
  10.1049/ip-opt:20000282} {\bibfield  {journal} {\bibinfo  {journal} {IEE
  Proceedings - Optoelectronics}\ }\textbf {\bibinfo {volume} {147}},\ \bibinfo
  {pages} {251--278} (\bibinfo {year} {2000})}\BibitemShut {NoStop}%
\bibitem [{\citenamefont {Keller}\ and\ \citenamefont
  {Tropper}(2006)}]{keller_passively_2006}%
  \BibitemOpen
  \bibfield  {author} {\bibinfo {author} {\bibfnamefont {U.}~\bibnamefont
  {Keller}}\ and\ \bibinfo {author} {\bibfnamefont {A.~C.}\ \bibnamefont
  {Tropper}},\ }\bibfield  {title} {\enquote {\bibinfo {title} {Passively
  modelocked surface-emitting semiconductor lasers},}\ }\href {\doibase
  10.1016/j.physrep.2006.03.004} {\bibfield  {journal} {\bibinfo  {journal}
  {Physics Reports}\ }\textbf {\bibinfo {volume} {429}},\ \bibinfo {pages}
  {67--120} (\bibinfo {year} {2006})}\BibitemShut {NoStop}%
\bibitem [{\citenamefont {Rafailov}, \citenamefont {Cataluna},\ and\
  \citenamefont {Sibbett}(2007)}]{rafailov_mode-locked_2007}%
  \BibitemOpen
  \bibfield  {author} {\bibinfo {author} {\bibfnamefont {E.~U.}\ \bibnamefont
  {Rafailov}}, \bibinfo {author} {\bibfnamefont {M.~A.}\ \bibnamefont
  {Cataluna}}, \ and\ \bibinfo {author} {\bibfnamefont {W.}~\bibnamefont
  {Sibbett}},\ }\bibfield  {title} {\enquote {\bibinfo {title} {Mode-locked
  quantum-dot lasers},}\ }\href
  {http://www.nature.com/nphoton/journal/v1/n7/abs/nphoton.2007.120.html}
  {\bibfield  {journal} {\bibinfo  {journal} {Nature photonics}\ }\textbf
  {\bibinfo {volume} {1}},\ \bibinfo {pages} {395--401} (\bibinfo {year}
  {2007})}\BibitemShut {NoStop}%
\bibitem [{\citenamefont {Hugi}\ \emph {et~al.}(2012)\citenamefont {Hugi},
  \citenamefont {Villares}, \citenamefont {Blaser}, \citenamefont {Liu},\ and\
  \citenamefont {Faist}}]{hugi_mid-infrared_2012}%
  \BibitemOpen
  \bibfield  {author} {\bibinfo {author} {\bibfnamefont {A.}~\bibnamefont
  {Hugi}}, \bibinfo {author} {\bibfnamefont {G.}~\bibnamefont {Villares}},
  \bibinfo {author} {\bibfnamefont {S.}~\bibnamefont {Blaser}}, \bibinfo
  {author} {\bibfnamefont {H.~C.}\ \bibnamefont {Liu}}, \ and\ \bibinfo
  {author} {\bibfnamefont {J.}~\bibnamefont {Faist}},\ }\bibfield  {title}
  {\enquote {\bibinfo {title} {Mid-infrared frequency comb based on a quantum
  cascade laser},}\ }\href {\doibase 10.1038/nature11620} {\bibfield  {journal}
  {\bibinfo  {journal} {Nature}\ }\textbf {\bibinfo {volume} {492}},\ \bibinfo
  {pages} {229--233} (\bibinfo {year} {2012})}\BibitemShut {NoStop}%
\bibitem [{\citenamefont {Palik}\ and\ \citenamefont
  {Ghosh}(1998)}]{palik_handbook_1998}%
  \BibitemOpen
  \bibinfo {editor} {\bibfnamefont {E.~D.}\ \bibnamefont {Palik}}\ and\
  \bibinfo {editor} {\bibfnamefont {G.}~\bibnamefont {Ghosh}},\ eds.,\
  \href@noop {} {\emph {\bibinfo {title} {Handbook of optical constants of
  solids}}}\ (\bibinfo  {publisher} {Academic Press},\ \bibinfo {address} {San
  Diego},\ \bibinfo {year} {1998})\BibitemShut {NoStop}%
\bibitem [{\citenamefont {Villares}\ \emph {et~al.}(2016)\citenamefont
  {Villares}, \citenamefont {Riedi}, \citenamefont {Wolf}, \citenamefont
  {Kazakov}, \citenamefont {S\"uess}, \citenamefont {Jouy}, \citenamefont
  {Beck},\ and\ \citenamefont {Faist}}]{villares_dispersion_2016}%
  \BibitemOpen
  \bibfield  {author} {\bibinfo {author} {\bibfnamefont {G.}~\bibnamefont
  {Villares}}, \bibinfo {author} {\bibfnamefont {S.}~\bibnamefont {Riedi}},
  \bibinfo {author} {\bibfnamefont {J.}~\bibnamefont {Wolf}}, \bibinfo {author}
  {\bibfnamefont {D.}~\bibnamefont {Kazakov}}, \bibinfo {author} {\bibfnamefont
  {M.~J.}\ \bibnamefont {S\"uess}}, \bibinfo {author} {\bibfnamefont
  {P.}~\bibnamefont {Jouy}}, \bibinfo {author} {\bibfnamefont {M.}~\bibnamefont
  {Beck}}, \ and\ \bibinfo {author} {\bibfnamefont {J.}~\bibnamefont {Faist}},\
  }\bibfield  {title} {\enquote {\bibinfo {title} {Dispersion engineering of
  quantum cascade laser frequency combs},}\ }\href {\doibase
  10.1364/OPTICA.3.000252} {\bibfield  {journal} {\bibinfo  {journal} {Optica}\
  }\textbf {\bibinfo {volume} {3}},\ \bibinfo {pages} {252} (\bibinfo {year}
  {2016})}\BibitemShut {NoStop}%
\bibitem [{\citenamefont {Kim}, \citenamefont {Choi},\ and\ \citenamefont
  {Delfyett}(2006)}]{kim_pulse_2006}%
  \BibitemOpen
  \bibfield  {author} {\bibinfo {author} {\bibfnamefont {J.}~\bibnamefont
  {Kim}}, \bibinfo {author} {\bibfnamefont {M.-T.}\ \bibnamefont {Choi}}, \
  and\ \bibinfo {author} {\bibfnamefont {P.~J.}\ \bibnamefont {Delfyett}},\
  }\bibfield  {title} {\enquote {\bibinfo {title} {Pulse generation and
  compression via ground and excited states from a grating coupled passively
  mode-locked quantum dot two-section diode laser},}\ }\href {\doibase
  10.1063/1.2410217} {\bibfield  {journal} {\bibinfo  {journal} {Applied
  Physics Letters}\ }\textbf {\bibinfo {volume} {89}},\ \bibinfo {pages}
  {261106} (\bibinfo {year} {2006})}\BibitemShut {NoStop}%
\bibitem [{\citenamefont {Diddams}(2010)}]{diddams_evolving_2010}%
  \BibitemOpen
  \bibfield  {author} {\bibinfo {author} {\bibfnamefont {S.~A.}\ \bibnamefont
  {Diddams}},\ }\bibfield  {title} {\enquote {\bibinfo {title} {The evolving
  optical frequency comb [{Invited}]},}\ }\href {\doibase
  10.1364/JOSAB.27.000B51} {\bibfield  {journal} {\bibinfo  {journal} {Journal
  of the Optical Society of America B}\ }\textbf {\bibinfo {volume} {27}},\
  \bibinfo {pages} {B51} (\bibinfo {year} {2010})}\BibitemShut {NoStop}%
\bibitem [{\citenamefont {Keilmann}, \citenamefont {Gohle},\ and\ \citenamefont
  {Holzwarth}(2004)}]{keilmann_time-domain_2004}%
  \BibitemOpen
  \bibfield  {author} {\bibinfo {author} {\bibfnamefont {F.}~\bibnamefont
  {Keilmann}}, \bibinfo {author} {\bibfnamefont {C.}~\bibnamefont {Gohle}}, \
  and\ \bibinfo {author} {\bibfnamefont {R.}~\bibnamefont {Holzwarth}},\
  }\bibfield  {title} {\enquote {\bibinfo {title} {Time-domain mid-infrared
  frequency-comb spectrometer},}\ }\href
  {https://www.osapublishing.org/abstract.cfm?uri=OL-29-13-1542} {\bibfield
  {journal} {\bibinfo  {journal} {Optics letters}\ }\textbf {\bibinfo {volume}
  {29}},\ \bibinfo {pages} {1542--1544} (\bibinfo {year} {2004})}\BibitemShut
  {NoStop}%
\bibitem [{\citenamefont {R\"osch}\ \emph {et~al.}(2014)\citenamefont
  {R\"osch}, \citenamefont {Scalari}, \citenamefont {Beck},\ and\ \citenamefont
  {Faist}}]{rosch_octave-spanning_2014}%
  \BibitemOpen
  \bibfield  {author} {\bibinfo {author} {\bibfnamefont {M.}~\bibnamefont
  {R\"osch}}, \bibinfo {author} {\bibfnamefont {G.}~\bibnamefont {Scalari}},
  \bibinfo {author} {\bibfnamefont {M.}~\bibnamefont {Beck}}, \ and\ \bibinfo
  {author} {\bibfnamefont {J.}~\bibnamefont {Faist}},\ }\bibfield  {title}
  {\enquote {\bibinfo {title} {Octave-spanning semiconductor laser},}\ }\href
  {\doibase 10.1038/nphoton.2014.279} {\bibfield  {journal} {\bibinfo
  {journal} {Nature Photonics}\ ,\ \bibinfo {pages} {42--47}} (\bibinfo {year}
  {2014})}\BibitemShut {NoStop}%
\bibitem [{\citenamefont {Burghoff}\ \emph {et~al.}(2014)\citenamefont
  {Burghoff}, \citenamefont {Kao}, \citenamefont {Han}, \citenamefont {Chan},
  \citenamefont {Cai}, \citenamefont {Yang}, \citenamefont {Hayton},
  \citenamefont {Gao}, \citenamefont {Reno},\ and\ \citenamefont
  {Hu}}]{burghoff_terahertz_2014}%
  \BibitemOpen
  \bibfield  {author} {\bibinfo {author} {\bibfnamefont {D.}~\bibnamefont
  {Burghoff}}, \bibinfo {author} {\bibfnamefont {T.-Y.}\ \bibnamefont {Kao}},
  \bibinfo {author} {\bibfnamefont {N.}~\bibnamefont {Han}}, \bibinfo {author}
  {\bibfnamefont {C.~W.~I.}\ \bibnamefont {Chan}}, \bibinfo {author}
  {\bibfnamefont {X.}~\bibnamefont {Cai}}, \bibinfo {author} {\bibfnamefont
  {Y.}~\bibnamefont {Yang}}, \bibinfo {author} {\bibfnamefont {D.~J.}\
  \bibnamefont {Hayton}}, \bibinfo {author} {\bibfnamefont {J.-R.}\
  \bibnamefont {Gao}}, \bibinfo {author} {\bibfnamefont {J.~L.}\ \bibnamefont
  {Reno}}, \ and\ \bibinfo {author} {\bibfnamefont {Q.}~\bibnamefont {Hu}},\
  }\bibfield  {title} {\enquote {\bibinfo {title} {Terahertz laser frequency
  combs},}\ }\href {\doibase 10.1038/nphoton.2014.85} {\bibfield  {journal}
  {\bibinfo  {journal} {Nature Photonics}\ }\textbf {\bibinfo {volume} {8}},\
  \bibinfo {pages} {462--467} (\bibinfo {year} {2014})}\BibitemShut {NoStop}%
\bibitem [{\citenamefont {Villares}\ \emph {et~al.}(2014)\citenamefont
  {Villares}, \citenamefont {Hugi}, \citenamefont {Blaser},\ and\ \citenamefont
  {Faist}}]{villares_dual-comb_2014}%
  \BibitemOpen
  \bibfield  {author} {\bibinfo {author} {\bibfnamefont {G.}~\bibnamefont
  {Villares}}, \bibinfo {author} {\bibfnamefont {A.}~\bibnamefont {Hugi}},
  \bibinfo {author} {\bibfnamefont {S.}~\bibnamefont {Blaser}}, \ and\ \bibinfo
  {author} {\bibfnamefont {J.}~\bibnamefont {Faist}},\ }\bibfield  {title}
  {\enquote {\bibinfo {title} {Dual-comb spectroscopy based on
  quantum-cascade-laser frequency combs},}\ }\href {\doibase
  10.1038/ncomms6192} {\bibfield  {journal} {\bibinfo  {journal} {Nature
  Communications}\ }\textbf {\bibinfo {volume} {5}},\ \bibinfo {pages} {5192}
  (\bibinfo {year} {2014})}\BibitemShut {NoStop}%
\bibitem [{\citenamefont {Yang}\ \emph {et~al.}(2016)\citenamefont {Yang},
  \citenamefont {Burghoff}, \citenamefont {Hayton}, \citenamefont {Gao},
  \citenamefont {Reno},\ and\ \citenamefont {Hu}}]{yang_terahertz_2016}%
  \BibitemOpen
  \bibfield  {author} {\bibinfo {author} {\bibfnamefont {Y.}~\bibnamefont
  {Yang}}, \bibinfo {author} {\bibfnamefont {D.}~\bibnamefont {Burghoff}},
  \bibinfo {author} {\bibfnamefont {D.~J.}\ \bibnamefont {Hayton}}, \bibinfo
  {author} {\bibfnamefont {J.-R.}\ \bibnamefont {Gao}}, \bibinfo {author}
  {\bibfnamefont {J.~L.}\ \bibnamefont {Reno}}, \ and\ \bibinfo {author}
  {\bibfnamefont {Q.}~\bibnamefont {Hu}},\ }\bibfield  {title} {\enquote
  {\bibinfo {title} {Terahertz multiheterodyne spectroscopy using laser
  frequency combs},}\ }\href {\doibase 10.1364/OPTICA.3.000499} {\bibfield
  {journal} {\bibinfo  {journal} {Optica}\ }\textbf {\bibinfo {volume} {3}},\
  \bibinfo {pages} {499} (\bibinfo {year} {2016})}\BibitemShut {NoStop}%
\bibitem [{\citenamefont {Westberg}, \citenamefont {Sterczewski},\ and\
  \citenamefont {Wysocki}(2017)}]{westberg_mid-infrared_2017}%
  \BibitemOpen
  \bibfield  {author} {\bibinfo {author} {\bibfnamefont {J.}~\bibnamefont
  {Westberg}}, \bibinfo {author} {\bibfnamefont {L.~A.}\ \bibnamefont
  {Sterczewski}}, \ and\ \bibinfo {author} {\bibfnamefont {G.}~\bibnamefont
  {Wysocki}},\ }\bibfield  {title} {\enquote {\bibinfo {title} {Mid-infrared
  multiheterodyne spectroscopy with phase-locked quantum cascade lasers},}\
  }\href {\doibase 10.1063/1.4979825} {\bibfield  {journal} {\bibinfo
  {journal} {Applied Physics Letters}\ }\textbf {\bibinfo {volume} {110}},\
  \bibinfo {pages} {141108} (\bibinfo {year} {2017})}\BibitemShut {NoStop}%
\bibitem [{\citenamefont {Lu}\ \emph {et~al.}(2017)\citenamefont {Lu},
  \citenamefont {Manna}, \citenamefont {Slivken}, \citenamefont {Wu},\ and\
  \citenamefont {Razeghi}}]{lu_dispersion_2017}%
  \BibitemOpen
  \bibfield  {author} {\bibinfo {author} {\bibfnamefont {Q.~Y.}\ \bibnamefont
  {Lu}}, \bibinfo {author} {\bibfnamefont {S.}~\bibnamefont {Manna}}, \bibinfo
  {author} {\bibfnamefont {S.}~\bibnamefont {Slivken}}, \bibinfo {author}
  {\bibfnamefont {D.~H.}\ \bibnamefont {Wu}}, \ and\ \bibinfo {author}
  {\bibfnamefont {M.}~\bibnamefont {Razeghi}},\ }\bibfield  {title} {\enquote
  {\bibinfo {title} {Dispersion compensated mid-infrared quantum cascade laser
  frequency comb with high power output},}\ }\href {\doibase 10.1063/1.4982673}
  {\bibfield  {journal} {\bibinfo  {journal} {AIP Advances}\ }\textbf {\bibinfo
  {volume} {7}},\ \bibinfo {pages} {045313} (\bibinfo {year}
  {2017})}\BibitemShut {NoStop}%
\bibitem [{\citenamefont {Riemensberger}\ \emph {et~al.}(2012)\citenamefont
  {Riemensberger}, \citenamefont {Hartinger}, \citenamefont {Herr},
  \citenamefont {Brasch}, \citenamefont {Holzwarth},\ and\ \citenamefont
  {Kippenberg}}]{riemensberger_dispersion_2012}%
  \BibitemOpen
  \bibfield  {author} {\bibinfo {author} {\bibfnamefont {J.}~\bibnamefont
  {Riemensberger}}, \bibinfo {author} {\bibfnamefont {K.}~\bibnamefont
  {Hartinger}}, \bibinfo {author} {\bibfnamefont {T.}~\bibnamefont {Herr}},
  \bibinfo {author} {\bibfnamefont {V.}~\bibnamefont {Brasch}}, \bibinfo
  {author} {\bibfnamefont {R.}~\bibnamefont {Holzwarth}}, \ and\ \bibinfo
  {author} {\bibfnamefont {T.~J.}\ \bibnamefont {Kippenberg}},\ }\bibfield
  {title} {\enquote {\bibinfo {title} {Dispersion engineering of thick high-{Q}
  silicon nitride ring-resonators via atomic layer deposition},}\ }\href
  {\doibase 10.1364/OE.20.027661} {\bibfield  {journal} {\bibinfo  {journal}
  {Optics Express}\ }\textbf {\bibinfo {volume} {20}},\ \bibinfo {pages}
  {27661} (\bibinfo {year} {2012})}\BibitemShut {NoStop}%
\bibitem [{\citenamefont {Okawachi}\ \emph {et~al.}(2014)\citenamefont
  {Okawachi}, \citenamefont {Lamont}, \citenamefont {Luke}, \citenamefont
  {Carvalho}, \citenamefont {Yu}, \citenamefont {Lipson},\ and\ \citenamefont
  {Gaeta}}]{okawachi_bandwidth_2014}%
  \BibitemOpen
  \bibfield  {author} {\bibinfo {author} {\bibfnamefont {Y.}~\bibnamefont
  {Okawachi}}, \bibinfo {author} {\bibfnamefont {M.~R.~E.}\ \bibnamefont
  {Lamont}}, \bibinfo {author} {\bibfnamefont {K.}~\bibnamefont {Luke}},
  \bibinfo {author} {\bibfnamefont {D.~O.}\ \bibnamefont {Carvalho}}, \bibinfo
  {author} {\bibfnamefont {M.}~\bibnamefont {Yu}}, \bibinfo {author}
  {\bibfnamefont {M.}~\bibnamefont {Lipson}}, \ and\ \bibinfo {author}
  {\bibfnamefont {A.~L.}\ \bibnamefont {Gaeta}},\ }\bibfield  {title} {\enquote
  {\bibinfo {title} {Bandwidth shaping of microresonator-based frequency combs
  via dispersion engineering},}\ }\href {\doibase 10.1364/OL.39.003535}
  {\bibfield  {journal} {\bibinfo  {journal} {Optics Letters}\ }\textbf
  {\bibinfo {volume} {39}},\ \bibinfo {pages} {3535} (\bibinfo {year}
  {2014})}\BibitemShut {NoStop}%
\bibitem [{\citenamefont {Peschel}, \citenamefont {Peschel},\ and\
  \citenamefont {Lederer}(1995)}]{peschel_compact_1995}%
  \BibitemOpen
  \bibfield  {author} {\bibinfo {author} {\bibfnamefont {U.}~\bibnamefont
  {Peschel}}, \bibinfo {author} {\bibfnamefont {T.}~\bibnamefont {Peschel}}, \
  and\ \bibinfo {author} {\bibfnamefont {F.}~\bibnamefont {Lederer}},\
  }\bibfield  {title} {\enquote {\bibinfo {title} {A compact device for highly
  efficient dispersion compensation in fiber transmission},}\ }\href {\doibase
  10.1063/1.114736} {\bibfield  {journal} {\bibinfo  {journal} {Applied Physics
  Letters}\ }\textbf {\bibinfo {volume} {67}},\ \bibinfo {pages} {2111}
  (\bibinfo {year} {1995})}\BibitemShut {NoStop}%
\bibitem [{\citenamefont {Bidaux}\ \emph
  {et~al.}(2017{\natexlab{a}})\citenamefont {Bidaux}, \citenamefont
  {Sergachev}, \citenamefont {Wuester}, \citenamefont {Maulini}, \citenamefont
  {Gresch}, \citenamefont {Bismuto}, \citenamefont {Blaser}, \citenamefont
  {Muller},\ and\ \citenamefont {Faist}}]{bidaux_plasmon-enhanced_2017}%
  \BibitemOpen
  \bibfield  {author} {\bibinfo {author} {\bibfnamefont {Y.}~\bibnamefont
  {Bidaux}}, \bibinfo {author} {\bibfnamefont {I.}~\bibnamefont {Sergachev}},
  \bibinfo {author} {\bibfnamefont {W.}~\bibnamefont {Wuester}}, \bibinfo
  {author} {\bibfnamefont {R.}~\bibnamefont {Maulini}}, \bibinfo {author}
  {\bibfnamefont {T.}~\bibnamefont {Gresch}}, \bibinfo {author} {\bibfnamefont
  {A.}~\bibnamefont {Bismuto}}, \bibinfo {author} {\bibfnamefont
  {S.}~\bibnamefont {Blaser}}, \bibinfo {author} {\bibfnamefont
  {A.}~\bibnamefont {Muller}}, \ and\ \bibinfo {author} {\bibfnamefont
  {J.}~\bibnamefont {Faist}},\ }\bibfield  {title} {\enquote {\bibinfo {title}
  {Plasmon-enhanced waveguide for dispersion compensation in mid-infrared
  quantum cascade laser frequency combs},}\ }\href {\doibase
  10.1364/OL.42.001604} {\bibfield  {journal} {\bibinfo  {journal} {Optics
  Letters}\ }\textbf {\bibinfo {volume} {42}},\ \bibinfo {pages} {1604}
  (\bibinfo {year} {2017}{\natexlab{a}})}\BibitemShut {NoStop}%
\bibitem [{\citenamefont {Jouy}\ \emph {et~al.}(2017)\citenamefont {Jouy},
  \citenamefont {Wolf}, \citenamefont {Bidaux}, \citenamefont {Allmendinger},
  \citenamefont {Mangold}, \citenamefont {Beck},\ and\ \citenamefont
  {Faist}}]{jouy_dual_2017}%
  \BibitemOpen
  \bibfield  {author} {\bibinfo {author} {\bibfnamefont {P.}~\bibnamefont
  {Jouy}}, \bibinfo {author} {\bibfnamefont {J.~M.}\ \bibnamefont {Wolf}},
  \bibinfo {author} {\bibfnamefont {Y.}~\bibnamefont {Bidaux}}, \bibinfo
  {author} {\bibfnamefont {P.}~\bibnamefont {Allmendinger}}, \bibinfo {author}
  {\bibfnamefont {M.}~\bibnamefont {Mangold}}, \bibinfo {author} {\bibfnamefont
  {M.}~\bibnamefont {Beck}}, \ and\ \bibinfo {author} {\bibfnamefont
  {J.}~\bibnamefont {Faist}},\ }\bibfield  {title} {\enquote {\bibinfo {title}
  {Dual comb operation of $\lambda$ 8.2 $\mu$m {Quantum} {Cascade} {Laser}
  frequency comb with 1 {W} optical power},}\ }\href@noop {} {\bibfield
  {journal} {\bibinfo  {journal} {Applied Physics Letters}\ }\textbf {\bibinfo
  {volume} {111}},\ \bibinfo {pages} {141102} (\bibinfo {year}
  {2017})}\BibitemShut {NoStop}%
\bibitem [{\citenamefont {Bidaux}\ \emph
  {et~al.}(2017{\natexlab{b}})\citenamefont {Bidaux}, \citenamefont {Fedorova},
  \citenamefont {Livshits}, \citenamefont {Rafailov},\ and\ \citenamefont
  {Faist}}]{bidaux_investigation_2017}%
  \BibitemOpen
  \bibfield  {author} {\bibinfo {author} {\bibfnamefont {Y.}~\bibnamefont
  {Bidaux}}, \bibinfo {author} {\bibfnamefont {K.~A.}\ \bibnamefont
  {Fedorova}}, \bibinfo {author} {\bibfnamefont {D.~A.}\ \bibnamefont
  {Livshits}}, \bibinfo {author} {\bibfnamefont {E.~U.}\ \bibnamefont
  {Rafailov}}, \ and\ \bibinfo {author} {\bibfnamefont {J.}~\bibnamefont
  {Faist}},\ }\bibfield  {title} {\enquote {\bibinfo {title} {Investigation of
  the chromatic dispersion in two-section {InAs}/{GaAs} quantum-dot lasers},}\
  }\href {\doibase 10.1109/LPT.2017.2768390} {\bibfield  {journal} {\bibinfo
  {journal} {IEEE Photonics Technology Letters}\ }\textbf {\bibinfo {volume}
  {29}},\ \bibinfo {pages} {2246 -- 2249} (\bibinfo {year}
  {2017}{\natexlab{b}})}\BibitemShut {NoStop}%
\bibitem [{\citenamefont {Yang}\ \emph {et~al.}(2008)\citenamefont {Yang},
  \citenamefont {L\"osch}, \citenamefont {Bronner}, \citenamefont {Hugger},
  \citenamefont {Fuchs}, \citenamefont {Aidam},\ and\ \citenamefont
  {Wagner}}]{yang_onhigh-peak-power_2008}%
  \BibitemOpen
  \bibfield  {author} {\bibinfo {author} {\bibfnamefont {Q.}~\bibnamefont
  {Yang}}, \bibinfo {author} {\bibfnamefont {R.}~\bibnamefont {L\"osch}},
  \bibinfo {author} {\bibfnamefont {W.}~\bibnamefont {Bronner}}, \bibinfo
  {author} {\bibfnamefont {S.}~\bibnamefont {Hugger}}, \bibinfo {author}
  {\bibfnamefont {F.}~\bibnamefont {Fuchs}}, \bibinfo {author} {\bibfnamefont
  {R.}~\bibnamefont {Aidam}}, \ and\ \bibinfo {author} {\bibfnamefont
  {J.}~\bibnamefont {Wagner}},\ }\bibfield  {title} {\enquote {\bibinfo {title}
  {{High}-peak-power strain-compensated {GaInAs}/{AlInAs} quantum cascade
  lasers (λ 4.6 $\mu$m) based on a slightly diagonal active region design},}\
  }\href {\doibase 10.1063/1.3054165} {\bibfield  {journal} {\bibinfo
  {journal} {Applied Physics Letters}\ }\textbf {\bibinfo {volume} {93}},\
  \bibinfo {pages} {251110} (\bibinfo {year} {2008})}\BibitemShut {NoStop}%
\bibitem [{\citenamefont {Bismuto}\ \emph {et~al.}(2011)\citenamefont
  {Bismuto}, \citenamefont {Terazzi}, \citenamefont {Beck},\ and\ \citenamefont
  {Faist}}]{bismuto_influence_2011}%
  \BibitemOpen
  \bibfield  {author} {\bibinfo {author} {\bibfnamefont {A.}~\bibnamefont
  {Bismuto}}, \bibinfo {author} {\bibfnamefont {R.}~\bibnamefont {Terazzi}},
  \bibinfo {author} {\bibfnamefont {M.}~\bibnamefont {Beck}}, \ and\ \bibinfo
  {author} {\bibfnamefont {J.}~\bibnamefont {Faist}},\ }\bibfield  {title}
  {\enquote {\bibinfo {title} {Influence of the growth temperature on the
  performances of strain-balanced quantum cascade lasers},}\ }\href {\doibase
  10.1063/1.3561754} {\bibfield  {journal} {\bibinfo  {journal} {Applied
  Physics Letters}\ }\textbf {\bibinfo {volume} {98}},\ \bibinfo {pages}
  {091105} (\bibinfo {year} {2011})}\BibitemShut {NoStop}%
\bibitem [{\citenamefont {Hofstetter}\ and\ \citenamefont
  {Faist}(1999)}]{hofstetter_measurement_1999}%
  \BibitemOpen
  \bibfield  {author} {\bibinfo {author} {\bibfnamefont {D.}~\bibnamefont
  {Hofstetter}}\ and\ \bibinfo {author} {\bibfnamefont {J.}~\bibnamefont
  {Faist}},\ }\bibfield  {title} {\enquote {\bibinfo {title} {Measurement of
  semiconductor laser gain and dispersion curves utilizing {Fourier} transforms
  of the emission spectra},}\ }\href
  {http://www.phys.ethz.ch/~mesoqc/old%20stuff/General/Publications/1999/99_PTL1372_DH.pdf}
  {\bibfield  {journal} {\bibinfo  {journal} {IEEE Photonics Technology
  Letters}\ }\textbf {\bibinfo {volume} {11}},\ \bibinfo {pages} {1372--1374}
  (\bibinfo {year} {1999})}\BibitemShut {NoStop}%
\end{thebibliography}
\end{document}

\title[]{Coupled waveguides for dispersion compensation in semiconductor lasers: Supplementary materials}
%\thanks{Footnote to title of article.}
\address{Institute for Quantum Electronics, ETH-Zurich, CH-8093 Zurich, Switzerland}
\author{Yves Bidaux}
\email{bidauxy@phys.ethz.ch} %% email address is required required
\author{Filippos Kapsalidis}
\author{Pierre Jouy}
\author{Mattias Beck}
\author{J\'er\^ome Faist}
\email{jfaist@ethz.ch} %% email address is required required
\date{\today}

\maketitle

\section*{Plasma frequency}

The plasma frequency~\cite{palik_handbook_1998} is shown as a function of the carrier density for InP and InGaAs in Supplementary Fig. \ref{plasma_frequency}. To fabricate a dispersion compensated QCL frequency comb based on a plasmon enhanced waveguide at 4.6 $\mu$m the doping concentration of the plasmonic layer should be increased to more than 10$^{20}$ cm$^{-1}$ which cannot be realised in practice. Therefore, for short infrared wavelengths this approach cannot be used.

\begin{figure}[h!]
\centering
\includegraphics[width=0.5\textwidth]{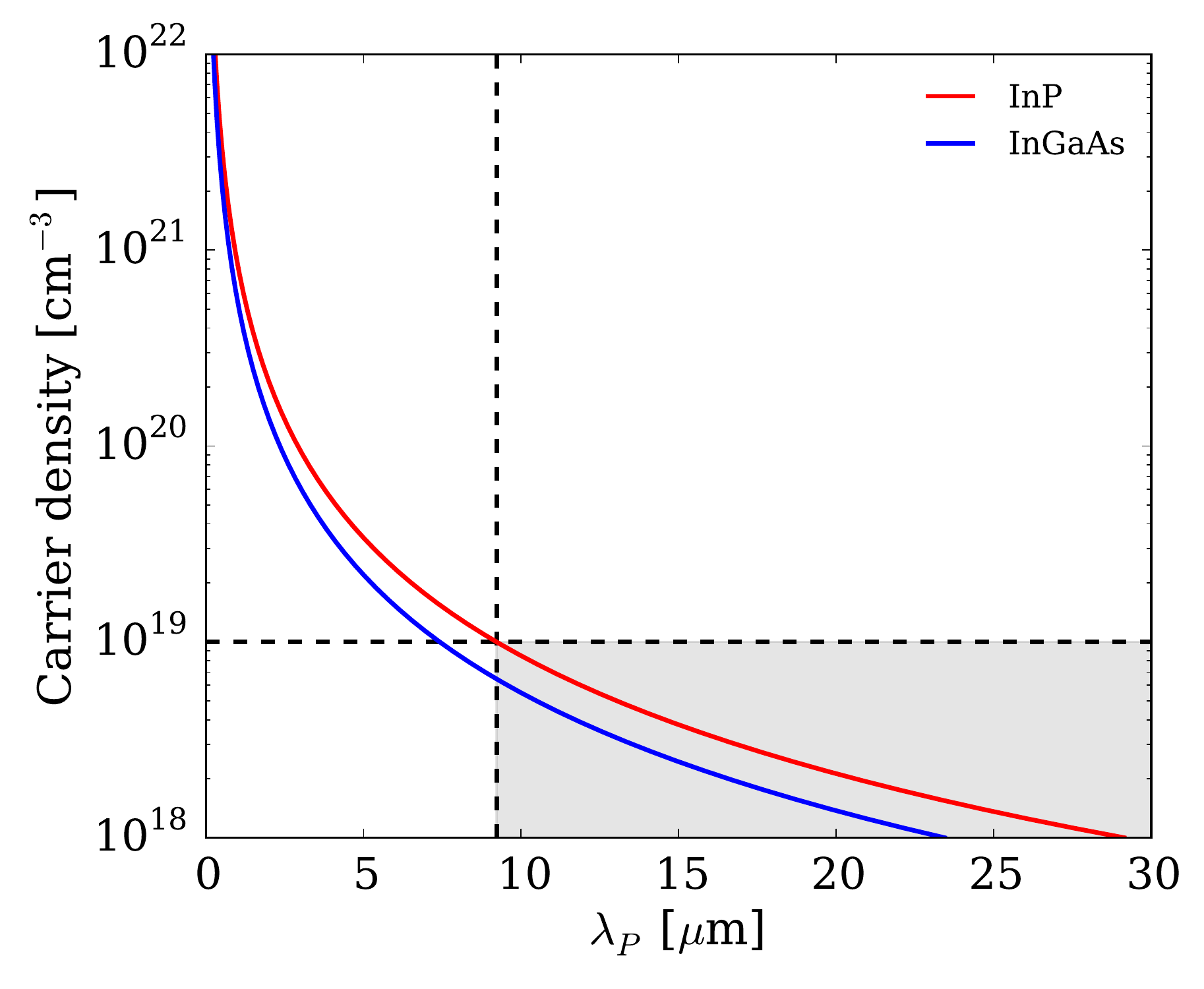}
\caption{\textbf{Plasma frequency:} Plasma frequency as a function of the carrier density for InP and InGaAs.}
\label{plasma_frequency}
\end{figure}

\section*{Coupled mode theory}

A model describing the dispersion of coupled waveguides and based on the coupled mode theory was proposed by Peschel et al.~\cite{peschel_compact_1995} The decoupled active region (i=1) and passive waveguide (i=2) modes have propagation constants $\beta_{1,2}$, group velocity $\nu_{1,2}$ and GVD D$_{1,2}$. The propagation vectors can be expanded around the frequency $\omega_0$ at which the propagation vectors are phase matched as
\begin{equation}
\beta_i\left( \omega \right) = \beta_0\left( \omega \right) + \left( \omega -\omega_0 \right) \frac{1}{\nu_i} + \frac{1}{2} \left( \omega -\omega_0 \right)^2 D_i.
\end{equation}
The modes of the coupled system are given by the eigensolutions of the matrix M:
\begin{equation}
\frac{da}{dz}=-Ma=-\left(\begin{array}{cc}
\delta & \kappa\\
\kappa & - \delta\end{array}\right)\cdot\left(\begin{array}{c}
a_{1}\\
a_{2}\end{array}\right),
\end{equation}
with $\delta = \left( \beta_1 - \beta_2 \right)/2$, $a_i$ the amplitudes of the modes and $\kappa$ the coupling constant between the two modes. The two optical modes couple to form two symmetric (+) and anti-symmetric (-) supermodes with propagation constants given by:
\begin{equation}
\beta_{\pm} = \frac{1}{2}\left( \beta_1 + \beta_2 \right) \pm \sqrt{\kappa^2 + \left( \beta_1 - \beta_2 \right)^2/4}.
\end{equation}
If the modes have the same GVD ($ GVD_0$), defining $\tilde{\omega} =\frac{\left( \omega -\omega_0\right)}{\delta \omega}$ where $\delta \omega = 2 \kappa  \abs{ \frac{1}{\nu_1} - \frac{1}{\nu_2} }  ^{-1}$ one gets
\begin{equation}
\beta^{'}_{+/-}  =  \frac{1}{2}\left( \frac{1}{\nu_1} + \frac{1}{\nu_2} \right) + \left( \omega -\omega_0\right) GVD_0 \pm \frac{\kappa \tilde{\omega}}{ \delta \omega} \left( \tilde{\omega}^2 +1 \right)^{1/2}
\end{equation}
and
\begin{equation}
\beta^{''}_{+/-}  =  GVD_0 \pm \frac{\kappa}{\delta\omega^2} \left[ \left( \tilde{\omega}^2 + 1 \right)^{-1/2}-  \tilde{\omega}^2\left( \tilde{\omega}^2 + 1 \right)^{-3/2}\right] \\
= GVD_0 \pm  \frac{\kappa}{\delta \omega^2 } \left( \tilde{\omega}^2 +1 \right)^{-3/2}
 \end{equation}
and the GVD of the supermodes is 
\begin{equation}
\label{peschel}
\text{GVD}_{+/-} = \text{GVD}_0 \pm\frac{1}{4 \kappa} \left( \frac{1}{\nu_1} - \frac{1}{\nu_2} \right) \left( \tilde{\omega}^2 +1 \right)^{-3/2}.
\end{equation}
The GVD induced by the coupling ($\kappa$) is positive for the symmetric mode and negative for the anti-symmetric mode. At $\omega_0$ the induced GVD is maximal and equals
\begin{equation}
\label{GVDpeak}
\text{GVD}^{peak}_{+/-} = \text{GVD}_{0} \pm \frac{1}{4 \kappa} \left( \frac{1}{\nu_1} - \frac{1}{\nu_2} \right) ^2.
\end{equation}

If the two modes now have a dissimilar GVD, the solution is given by
\begin{multline}
\text{GVD}_{+/-} = \frac{1}{2} \left( D_1 + D_2 \right)\\
\pm \frac{-\kappa^2 \tilde{\omega}^2 \left[ \left( \nu_1 - \nu_2 \right)^2 + 2\left( D_1 - D_2 \right) \nu_1^2  \nu_2^2 \kappa \tilde{\omega}\right]^2 \left[ \left( \nu_1 - \nu_2 \right)^2 + \left( D_1 - D_2 \right) \nu_1^2  \nu_2^2 \kappa \tilde{\omega}\right]}{\nu_1^2  \nu_2^2 \left( \nu_1 - \nu_2 \right)^6 \left(1 + \frac{\tilde{\omega}^2}{\left( \nu_1 - \nu_2 \right)^4} \left[ \left( \nu_1 - \nu_2 \right)^2 + \left( D_1 - D_2 \right) \nu_1^2  \nu_2^2 \kappa \tilde{\omega}\right]^2 \right)^{3/2} } \\
\pm  \frac{\left[ \left( \nu_1 - \nu_2 \right)^2 + \left( D_1 - D_2 \right) \nu_1^2  \nu_2^2 \kappa \tilde{\omega}\right]^2}{\nu_1^2  \nu_2^2 \left( \nu_1 - \nu_2 \right)^2 \kappa \sqrt{1 + \frac{\tilde{\omega}^2}{\left( \nu_1 - \nu_2 \right)^4} \left[ \left( \nu_1 - \nu_2 \right)^2 + \left( D_1 - D_2 \right) \nu_1^2  \nu_2^2 \kappa \tilde{\omega}\right]^2}} \\
\pm  \frac{2 \left( D_1 - D_2 \right) \tilde{\omega} \left[ \left( \nu_1 - \nu_2 \right)^2 + \left( D_1 - D_2 \right) \nu_1^2  \nu_2^2 \kappa \tilde{\omega}\right]}{\left( \nu_1 - \nu_2 \right)^2 \kappa \sqrt{1 + \frac{\tilde{\omega}^2}{\left( \nu_1 - \nu_2 \right)^4} \left[ \left( \nu_1 - \nu_2 \right)^2 + \left( D_1 - D_2 \right) \nu_1^2  \nu_2^2 \kappa \tilde{\omega}\right]^2}}.
\end{multline}

One can express the full solution under the form of equation \ref{peschel} by reorganizing the terms. One gets
\begin{equation}
\text{GVD}_{+/-} = \text{GVD}_0 \pm\frac{1}{4 \kappa} \left( \frac{1}{\nu_1} - \frac{1}{\nu_2} \right) \left( C_1\cdot \tilde{\omega}^2 +1 \right)^{-3/2}\cdot C_2,
\end{equation}
where the functions $C_1$ and $C_2$ contains the terms that depend on D$_1$-D$_2$:
\begin{equation}
C_1 = \frac{B^2}{\left( \nu_1 -\nu_2 \right)^4},
\end{equation}
\begin{equation}
C_2 = \frac{4 Z B+B^2+Z^2+2\tilde{\omega}^2 Z B^3}{\left( \nu_1 -\nu_2 \right)^4},
\end{equation}

with

\begin{equation}
B = Z +\left( \nu_1 -\nu_2 \right)^2
\end{equation}
and
\begin{equation}
Z = \left( D_1 - D_2 \right) \kappa \tilde{\omega} \nu_1^2 \nu_2^2.
\end{equation}
One can expand $C_1$ and $C_2$ around $\omega = \omega_0$ to the first order and get
\begin{equation}
C_1 \simeq 1,
\end{equation}
\begin{equation}
C_2 \simeq 1 + \alpha \tilde{\omega}.
\end{equation}
Finally, we obtain
\begin{equation}
\text{GVD}_{+/-} \simeq \text{GVD}_0 \pm\frac{1}{4 \kappa} \left( \frac{1}{\nu_1} - \frac{1}{\nu_2} \right) \left(\tilde{\omega}^2 +1 \right)^{-3/2}\left( 1 + \alpha \tilde{\omega} \right).
\end{equation}

The introduction of a GVD difference between the two waveguides has the effect to shift the GVD peaks away from $\omega_0$ by the introduction of an asymmetric factor which is given by
\begin{equation}
\alpha = \frac{6 \kappa}{\delta \omega^2} \left(D_1 -D_2 \right).
\end{equation}

\section*{Optical frequency combs characterization}

\begin{figure*}[h!]
\centering
\includegraphics[width=1\textwidth]{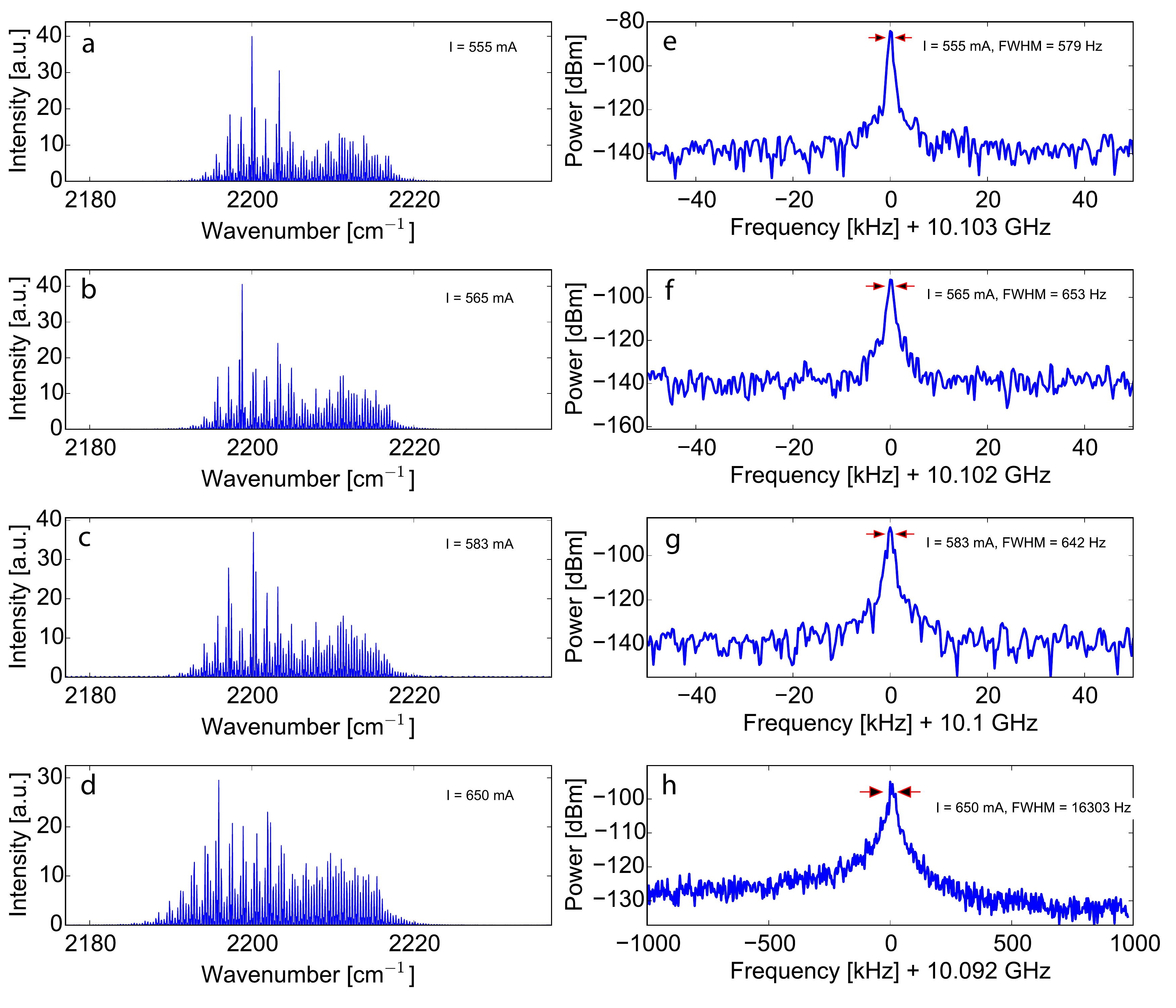}
\caption{\textbf{Reference laser characterization. a-d}, Optical spectra measured for the reference device at a heatsink temperature of -15$^{\circ}$C for a current of 555, 565, 583 and 650 mA. \textbf{e-h} Corresponding intermode beatnotes measured using a spectrum analyser with a RBW of 500 Hz.}
\label{EV2139A}
\end{figure*}

Supplementary Fig.~\ref{EV2139A} shows the optical spectra measured from the reference device for a heatsink temperature of -15$^{\circ}$C for a current of 555, 565, 583 and 650 mA as well as the corresponding intermode beatnotes measured using a spectrum analyser with a RBW of 500 Hz.  We choose this resolution bandwidth as a good compromise between resolution and acquisition time, to limit the influence of the drift during the measurement. Note that the results measured in ~\citet{hugi_mid-infrared_2012} were obtained using a real-time spectrum analyser, not a scanning instrument used here. At high currents the intermode beatnote is broad. Moreover, no multiheterodyne beating spectra could be measured from the device. This suggests that the reference device is not operating as an optical frequency comb.

\begin{figure*}[h!]
\centering
\includegraphics[width=1\textwidth]{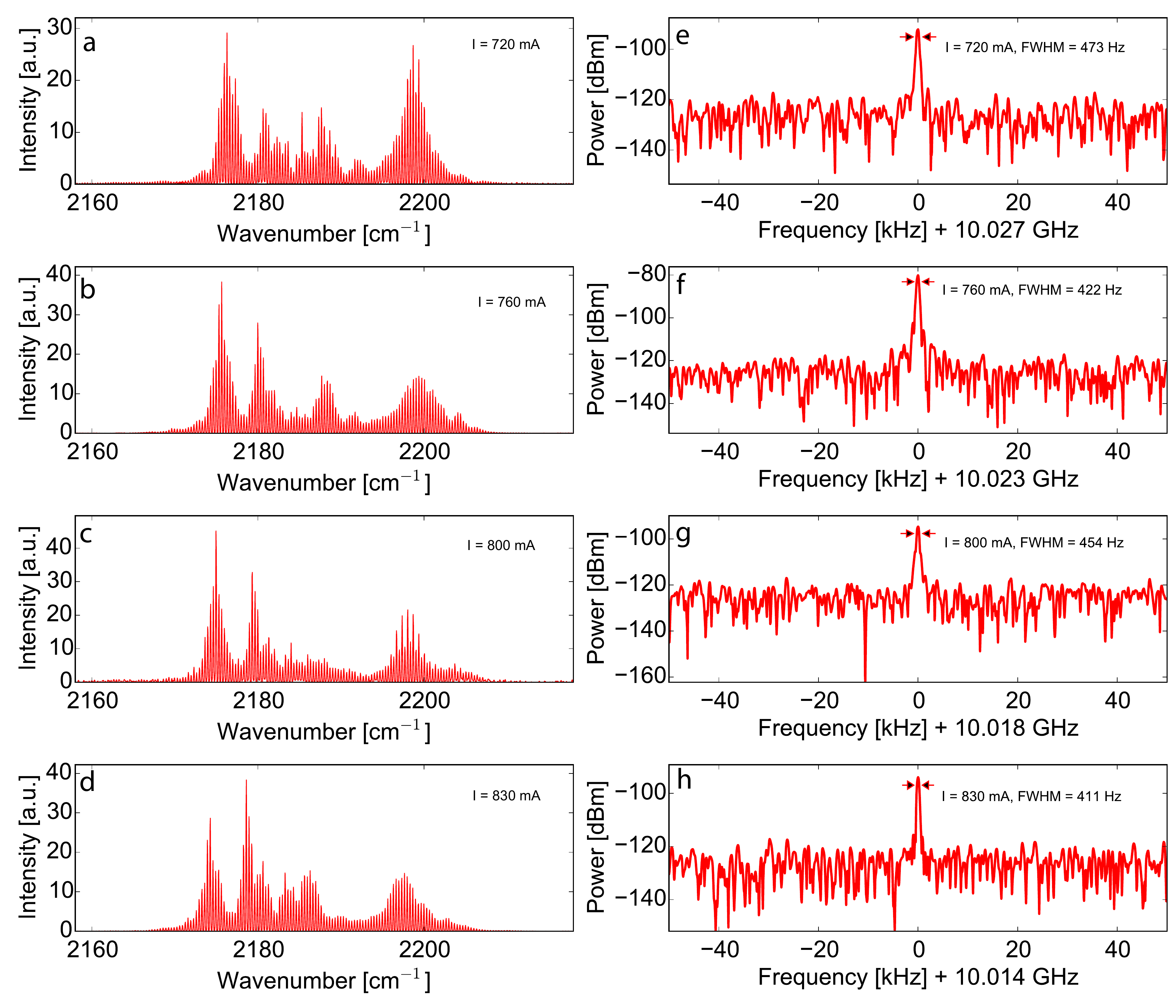}
\caption{\textbf{Dual waveguide laser characterization. a-d}, Optical spectra measured from the coupled waveguide device at a heatsink temperature of -15$^{\circ}$C for a current of 720, 760, 800 and 830 mA. \textbf{e-h} Corresponding intermode beatnotes measured using a spectrum analyser with a RBW of 500 Hz.}
\label{EV1527D}
\end{figure*}

Supplementary Fig.~\ref{EV1527D} shows the optical spectra measured from the coupled waveguide device for a heatsink temperature of -15$^{\circ}$C for a current of 720, 760, 800 and 830 mA as well as the corresponding intermode beatnotes measured using a spectrum analyser with a RBW of 500 Hz.

\clearpage

\section*{Dual waveguide for quantum-dot lasers}

We propose a dual waveguide for GaAs/AlGaAs quantum-dot lasers. The proposed core waveguide is composed of a 643 nm thick layer Al$_{0.15}$Ga$_{0.85}$As. The core waveguide is placed in between two Al$_{0.35}$Ga$_{0.65}$As 1520 nm cladding layers (Si doped n=10$^{17}$ cm$^{-3}$ and C doped p=10$^{18}$ cm$^{-3}$). A 360 nm passive GaAs waveguide is placed in the upper cladding layer at a distance d=700 nm from the primary core waveguide. The vertical refractive index profile in the device along the growth axis is displayed in Supplementary Fig.~\ref{fig:qdot_graph}a. The computed electric field profile of the anti-symmetric (blue line) and symmetric (red line) modes are also displayed, while the full two dimensional intensity profiles of these modes are reported in Supplementary Fig.~\ref{fig:qdot_graph}b and c.

In Supplementary Fig.~\ref{fig:qdot_gvd}a, the dispersion curves of the two uncoupled modes are reported with dashed lines while the dispersion of the coupled modes are reported with full lines. Our model predicts a GVD close to zero at the laser wavelength for the anti-symmetric mode (see Supplementary Fig.~\ref{fig:qdot_gvd}b). The anti-symmetric mode has a higher overlap factor with the active region while the symmetric mode is localized primarily in the GaAs waveguide. In that case, mode selection results from the overlap with the active region which is larger for the anti-symmetric mode (see Supplementary Fig.~\ref{fig:qdot_gvd}c).

\begin{figure*}[h]
\centering
\includegraphics[width=1\textwidth]{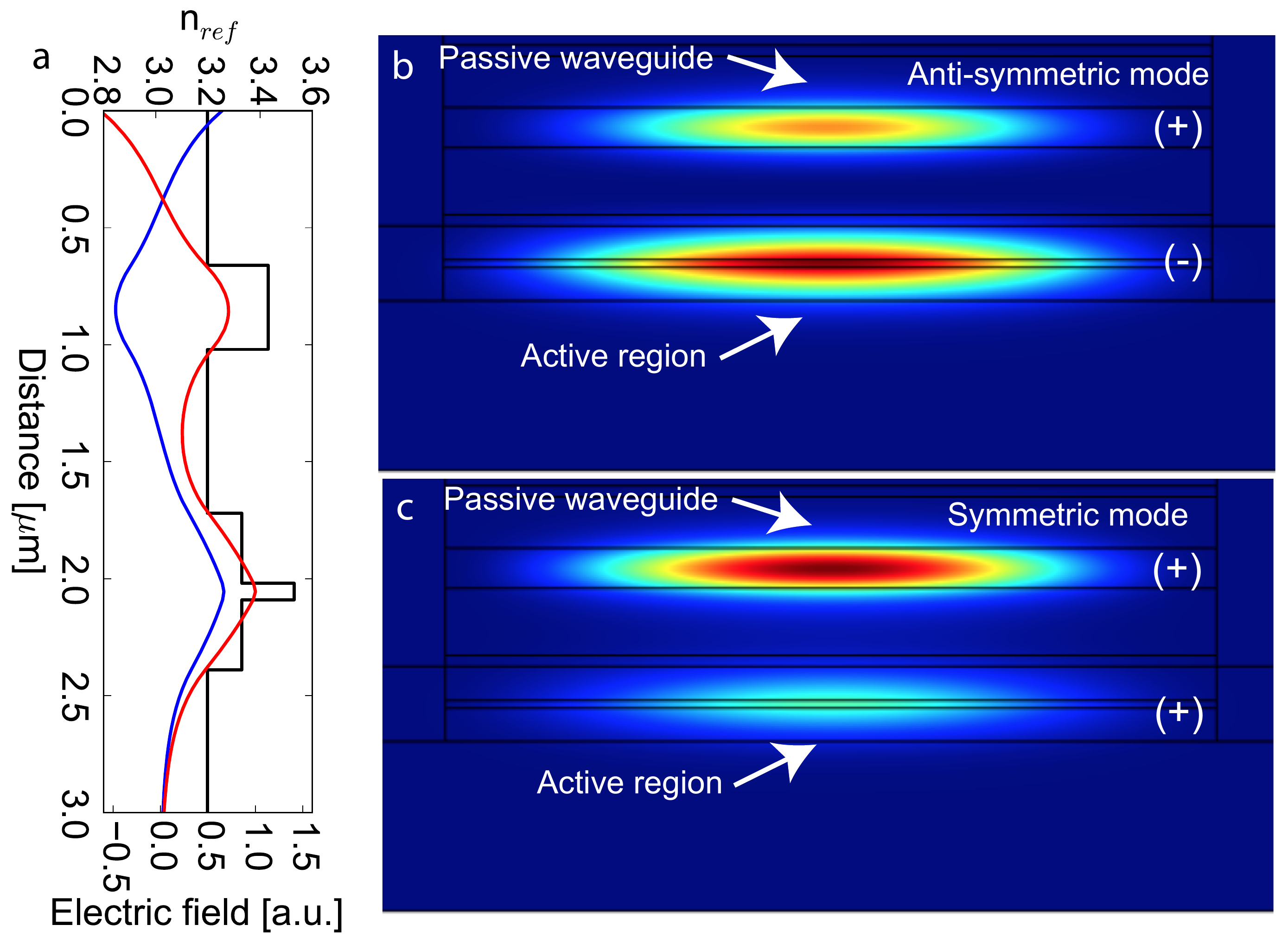}
\caption{\textbf{Device cross-section. a.} Cut in the vertical plan of the respective refractive index profile and electric field profile of the anti-symmetric (blue line) and symmetric (red line) modes. \textbf{b}, Intensity profile of the anti-symmetric mode at 8600 cm$^{-1}$. \textbf{c}, Intensity profile of the symmetric mode at 8600 cm$^{-1}$.}
\label{fig:qdot_graph}
\end{figure*}

\begin{figure*}[h!]
\centering
\includegraphics[width=0.75\textwidth]{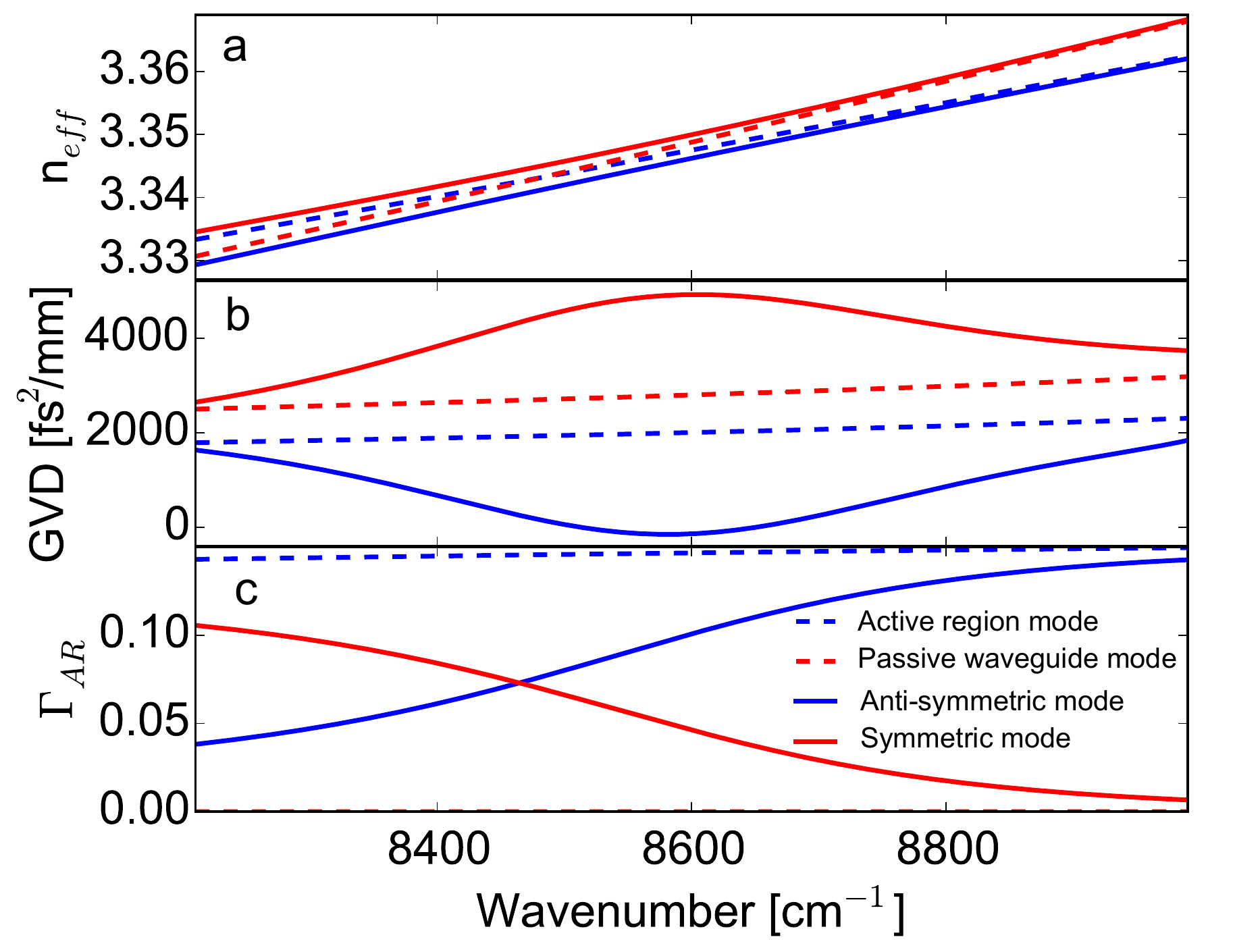}
\caption{\textbf{Simulation of the dispersion. a}, Spectrally resolved dispersion and \textbf{b}, GVD of the optical modes. \textbf{c}, Overlap factor with the active region.}
\label{fig:qdot_gvd}
\end{figure*}

\clearpage